\documentclass[11pt,a4paper]{article}
\pdfoutput=1
\usepackage{jheppub}
\usepackage[utf8]{inputenc}

\usepackage{color}
\usepackage{amsmath}
\usepackage{pifont}
\usepackage{bbold}

\usepackage{bbm}
\usepackage{verbatim}   
\usepackage{subfigure}
\usepackage{acronym}

\usepackage{amsfonts}
\usepackage{amssymb}
\usepackage{mathrsfs}
\usepackage{graphicx}
\usepackage{multirow}
\usepackage{slashed}
\usepackage{float}

\usepackage{caption}
\usepackage{tikz}
\usetikzlibrary{decorations.pathreplacing}
\newcommand{\diff}[2]{\frac{\mathrm{d} #1 }{\mathrm{d} #2 }}
\newcommand{\ddiff}[2]{\frac{\mathrm{d}^2 #1 }{\mathrm{d} #2^2 }}
\newcommand{\tbh}[0]{T_\mathrm{BH}}
\renewcommand{\Im}{\mathrm{Im}\,}

\title{Hawking Radiation of Extended Objects}

\author[1]{George Johnson}
\author[1]{John March-Russell}
\emailAdd{george.johnson@physics.ox.ac.uk}
\emailAdd{jmr@thphys.ox.ac.uk}

\affiliation[1]{Rudolf Peierls Centre for Theoretical Physics, University of Oxford, OX1 3NP, United Kingdom}

\abstract{We compute the effects on the temperature and precise spectrum of Hawking radiation from a Schwarzschild black hole when the emitted object is taken to be spatially extended. We find that in the low-momentum regime, the power emitted is exponentially suppressed for sufficiently large radiated objects, or sufficiently small black holes, though the temperature of emission is unchanged. We numerically determine the magnitude of this suppression as a function of the size and mass of the object and the black hole, and discuss the implications for various extended objects in nature.
}

\begin{document}

\maketitle

\section{Introduction}

Since Hawking's original discovery \cite{Hawking1,Hawking2} in 1974 that black holes emit radiation with a temperature inversely proportional to their mass, there has been a wealth of research into the precise spectrum of this radiation. In particular, there have been detailed studies of the rate of emission of particles of different masses and spins, for black holes charged and uncharged, spinning and static, and in different dimensions.

Several different derivations of the Hawking temperature exist in the literature; whilst the original analysis considers the scattering of wavepackets off the horizon towards future null infinity, Hawking and Gibbons, for instance, approached the topic from the perspective of the Euclidean path integral for the gravitational field \cite{Gibbons}. In 1999, Wilczek and Parikh showed that Hawking radiation can be viewed as a tunnelling process \cite{Wilczek}, in which particles pass through the contracting horizon of the black hole, lending formal justification to the most intuitive picture we have of black hole evaporation.

All such studies, however, assume the emitted particles are pointlike. In nature, of course, many particles are extended objects. Bound states of QCD, for instance, have finite size. Solitons, such as monopoles in gauge theories or Q-balls in scalar field theories, are not pointlike either. Even non-composite particles, such as fundamental strings or even black holes themselves, can have finite extent.

In this paper we study the effects of the extended nature of such objects on the temperature of the black hole and the precise spectrum of emission. The finite size of these objects modifies their interaction with the gravitational field; in particular, they are subject to tidal forces. We expect that these forces may alter the process of tunnelling through the horizon, as well as the subsequent evolution of the particles in the black hole background. For small black holes, such as the primordial black holes speculated to have existed in the early universe \cite{primordial}, these tidal forces should be large enough as to modify the spectrum of emission. In this paper we restrict attention to uncharged and nonrotating black holes, and consider only spinless emitted particles.

Deviation from the pure black-body spectrum predicted by Hawking's original calculation can be described by frequency-dependent \textit{greybody factors}. They arise due to the interaction of the emitted particles with the background gravitational field, after having been produced outside the horizon. These greybody factors are one of several effects that modify the pattern of radiation from that of a purely Planckian spectrum, whose supposed absence of correlations is the source of the black hole information puzzle \cite{thermality}.

This paper is divided into five sections. In Section 2, we discuss the general theory that determines the motion of extended objects in curved spacetimes. In particular, we establish an expansion of the action in powers of the curvature of the spacetime, that we can truncate to lowest order. In Section 3, we compute the temperature of a black-body that radiates extended objects, and show that it coincides with the usual Hawking temperature. In Section 4 we compute the greybody factors associated with emission of such particles, at least in the long wavelength regime. In this regime, only the $s$-wave solution is relevant. Finally, in Section 5, we discuss the results, and mention the implications for various finite size particles we observe, or hypothesise to exist, in nature.

We take $c=\hbar=1$ throughout. We write $M$ for the length $GM = M/M_P^2$.

\section{Theory of Extended Objects}

We will study the dynamics of extended objects in a background gravitational field  by approximating them as point-particles with some \textit{effective action} that is a modification of the typical relativistic point-particle action. All the effects of tidal forces on the body can be captured in a systematic fashion through terms coupling to the curvature of the metric. If we assume the extended object is spherical, the problem of writing down terms compatible with rotational symmetry, worldline reparametrization invariance and general coordinate invariance is straightforward \cite{houches}. The first few terms in such an effective action, linear in the curvature, are 
\begin{equation}
S = -m \int \mathrm{d}\tau - c_1 \int\mathrm{d}\tau \, R - c_2 \int \mathrm{d}\tau \, R_{\mu \nu} \dot{x}^\mu \dot{x}^\nu + \cdots
\end{equation}
where $c_1$, $c_2$ are constants, $R_{\mu \nu}$ is the Ricci tensor, and $R$ the Ricci scalar. For Schwarzschild spacetime, which will be the case we consider in this paper, $R$ and $R_{\mu \nu}$ are zero. The lowest-order non-vanishing terms in the action are constructed from the Riemann tensor, and involve two powers of the curvature. We will employ this effective action:
\begin{equation}
S = -\int \mathrm{d}\tau \Big(m + c \,E_{\mu \nu}E^{\mu \nu} + c'\,B_{\mu \nu}B^{\mu \nu}\Big) \,,
\end{equation}
where
\begin{align}
E_{\mu \nu} &= R_{\mu \alpha \nu \beta} \dot{x}^\alpha \dot{x}^\beta \,, \\
B_{\mu \nu} &= \epsilon_{\mu \alpha \beta \gamma} R^{\alpha \beta}{}_{\delta \nu} \dot{x}^\gamma \dot{x}^\delta \,,
\end{align}
and $c,c'$ are assumed to depend on the size of the object. In general the action depends on four powers of the particle velocity. Assuming purely radial motion, we find for Schwarzschild spacetime that $B_{\mu \nu}=0$\footnote{We emphasise that for more general motion, $B_{\mu \nu}$ is not zero.}, and the effective action can be written
\begin{equation}
S = -\int \mathrm{d} \tau \left(m + c \frac{6M^2}{r^6}\left((1-2M/r)\dot{t}^2 - \dot{r}^2 - 2 \dot{t}\dot{r} \sqrt{2M/r}\right)^2 \right) \,.
\end{equation}
We note that such objects no longer move on geodesics; the deviation from geodesic motion is due to the aforementioned tidal forces.

\section{Hawking Temperature for Extended Objects}
In this section we study the semi-classical tunnelling of extended objects through the black hole horizon, in line with the analysis of \cite{Wilczek}. See also \cite{selfinteraction,secret,energycons,tunnellingmethods} for further discussion. The rate of emission $\Gamma$ will have exponential part given by
\begin{equation}
\Gamma \sim \exp(-2\, \Im S) \,,
\end{equation}
where $S$ is the tunnelling action. According to the Planck law, particles with frequency $\omega$ should be radiated at a rate with exponential dependence $\Gamma \sim \exp(-\omega/T)$. We can hence read off the temperature at which the black hole radiates according to
\begin{equation}
\tbh{} = \frac{\omega}{2 \,\Im S} \,.
\end{equation}

We will study the process in Painlev\'{e}-Gullstrand coordinates. The time coordinate $t$, just as for Schwarzschild time, corresponds to the time measured by a stationary observer at infinity. Ignoring angular directions, the metric is
\begin{equation}
\mathrm{d}s^2 = -\left(1-\frac{2M}{r}\right)\mathrm{d}t^2 + 2 \sqrt{\frac{2M}{r}} \mathrm{d}t\, \mathrm{d}r + \mathrm{d}r^2 \,,
\end{equation}
where $M$ is the mass of the black hole. 

We will firstly present a calculation of the imaginary part of the action for a massless point-particle tunnelling through the horizon of Schwarzschild spacetime in the spirit of \cite{Wilczek}. We will then repeat the calculation for a massive particle, and finally present the calculation for extended objects. In all three cases we find that the temperature of the black hole is given by the usual Hawking formula $\tbh{}=1/8 \pi M$. We can argue on simple physical grounds that this must be the case. In particular, if a given black hole is in thermal equilibrium with a bath of particles of two different species, then by the zeroth law of thermodynamics, those two baths of particles must be in equilibrium with each other, and hence have the same temperature. It is nevertheless useful to see how the Hawking temperature falls explicitly out of the calculation in each case.

We can write the action for a particle moving freely in a curved background as
\begin{equation}
\label{action}
S = \int p_\mu \, \mathrm{d}x^\mu \qquad \text{with} \qquad p_\mu = g_{\mu \nu}\diff{x^\nu}{\sigma} \,,
\end{equation}
where $\sigma$ is an affine parameter along the worldline of the particle, chosen so that $p^\mu$ coincides with the physical 4-momentum of the particle. For a massive particle, this requires that $\mathrm{d}\sigma = \mathrm{d}\tau/m$, with $\tau$ the proper time.

\subsection{Case $m=0$}   

The radial dynamics of massless particles in Schwarzschild spacetime are determined by the equations
\begin{align}
\label{reqn}
\left(1- \frac{2M}{r}\right)\dot{t}^2 - 2\sqrt{\frac{2M}{r}}\dot{r}\dot{t} - \dot{r}^2 = 0  \,, \\
\label{teqn}
\left(1- \frac{2M}{r}\right)\dot{t} - \sqrt{\frac{2M}{r}}\dot{r} = \omega \,.
\end{align}
The second equation is the geodesic equation corresponding to the time-independence of the metric; in terms of the momentum defined in Eq. \eqref{action}, it can be written $p_t = \omega$, and so $\omega$ has the interpretation of the energy of the particle as measured at infinity. The first equation can be factorised to yield, for an outgoing particle, the equation
\begin{equation}
\diff{r}{t} = 1 - \sqrt{\frac{2M}{r}} \,.
\end{equation}
We then have
\begin{align}
\Im S &=  \Im \int \omega \, \mathrm{d}t+ \Im \int p_r \, \mathrm{d}r \\
&= \Im \int \left( \sqrt{\frac{2M}{r}} \dot{t} + \dot{r}\right) \mathrm{d}r \\
&= \Im \int \dot{t}\left( \sqrt{\frac{2M}{r}}  + \diff{r}{t}\right) \mathrm{d}r \\
&= \Im \int \dot{t}\,\mathrm{d}r \,.
\end{align}
Using Eqs. \eqref{reqn} and \eqref{teqn} we can write $\dot{t}$ in terms of $\omega$ to yield:
\begin{equation}
\Im S = \Im \int \frac{\omega}{1-\sqrt{2M/r}}\,\mathrm{d}r \,.
\end{equation}
The integrand has a pole at $r=2M$, the horizon. Choosing the prescription to integrate clockwise around this pole (into the upper-half complex-$r$ plane), we find
\begin{equation}
\Im S = 4 \pi M \omega \,,
\end{equation}
giving $\tbh{}=1/8 \pi M$, as expected.

\subsection{Case $m >0$}
For massive point-particles, the dynamics are governed by
\begin{align}
\label{reqnm}
\left(1- \frac{2M}{r}\right)\dot{t}^2 - 2\sqrt{\frac{2M}{r}}\dot{r}\dot{t} - \dot{r}^2 = 1 \,,\\
\label{teqnm}
\left(1- \frac{2M}{r}\right)\dot{t} - \sqrt{\frac{2M}{r}}\dot{r} = \omega \,.
\end{align}
Using these equations to solve for $p_r$ is somewhat more complicated than before, but essentially the same. Taking Eq. \eqref{teqnm} and substituting for $\dot{r}$ into Eq. \eqref{reqnm} yields a quadratic equation in $\dot{t}$ which we can solve:
\begin{equation}
\dot{t}^2 \left(\frac{r}{2M}-1\right) - \dot{t}\left(\frac{\omega r}{M}\right) + \left(\frac{r\omega^2}{2M} +1\right) = 0 \,.
\end{equation}
We find
\begin{align}
\left(1-\frac{2M}{r}\right) \dot{t} &= \omega \pm \sqrt{2M/r}\sqrt{\omega^2 - 1 + 2M/r} \,, \\
\dot{r} &= \pm\sqrt{\omega^2 - 1 +2M/r} \,.
\end{align}
From here we compute the imaginary part of the action thus:
\begin{align}
\Im S &= \Im \int p_r \, \mathrm{d}r\\
&= \Im \int \left( \sqrt{\frac{2M}{r}} \dot{t} + \dot{r}\right) \mathrm{d}r \\
&= \Im \int \left(\frac{\dot{r} + \omega\sqrt{2M/r}}{1-2M/r} + \dot{r}\right) \mathrm{d}r \,.
\end{align}
Taking the positive sign for $\dot{r}$ and substituting yields
\begin{equation}
\Im S = \Im \int \left(\frac{\sqrt{\omega^2 - 1 + 2M/r} + \omega\sqrt{2M/r}}{1-2M/r}\right) \mathrm{d}r \,.
\end{equation}
Finally we integrate around the pole at $r=2M$ as before. And as before, we find
\begin{equation}
\Im S = 4 \pi M \omega \,.
\end{equation}

\subsection{Case $c>0$}
In order to calculate the tunnelling rate for particles of finite size, we first need to determine the modifications to the equations of motion that result from the additional term in the action, before substituting these back into the action. Matters are simplified greatly for radial motion; using the standard normalisation of the 4-velocity, the action is simply
\begin{equation}
S = -\int \mathrm{d} \tau \left(m + c \frac{6M^2}{r^6} \right) 
\equiv -m\int \mathrm{d} \tau \, f(r) \,.
\label{simple}
\end{equation} 
As per usual, we can use the 4-velocity normalisation in place of the equation of motion for $r(\tau)$, leaving us with only one equation to determine. The $t$-independence of the effective action makes this relatively simple (cf. \eqref{teqn}):
\begin{equation}
\omega = \left((1-2M/r)\dot{t} - \sqrt{2M/r}\dot{r}\right)f(r) \,.
\end{equation}
The manipulations of the previous section proceed here as before, but with $\omega$ replaced everywhere by $\omega/f(r)$. We thus compute 
\begin{equation}
\Im S = \Im \int f(r)\, p_t \, \mathrm{d}t + \Im \int f(r)\, p_r \, \mathrm{d}r \,.
\end{equation}
The first term has no imaginary part. The second term is given by
\begin{equation}
\Im \int f(r) \,p_r \, \mathrm{d}r =
\Im \int f(r)\left(\frac{\sqrt{(\omega/f)^2 - 1 + 2M/r} + (\omega/f)\sqrt{2M/r}}{1-2M/r}\right) \mathrm{d}r \,.
\end{equation}
At the pole, the factors of $f(2M)$ cancel, yielding
\begin{equation}
\Im S = 4 \pi M \omega \,,
\end{equation}
as before.

\section{Greybody Factors}

We have found that a Schwarzschild black hole with mass $M$ radiates extended objects at the Hawking temperature $\tbh{}$. We next wish to find the exact rate at which such objects are radiated. For a perfect black-body emitting massless particles, the power received per solid angle subtended per unit frequency is given by the Planck law:
\begin{equation}
P(\omega) = A\frac{\omega^3}{8 \pi^3}\frac{1}{\exp(\omega/k_B T)-1} \,,
\end{equation}
where $A$ is the projected area of the black-body. For a non-gravitating sphere of radius $r$, we would have $A=\pi r^2$. For a black-body which is not perfect, this relation is modified to
\begin{equation}
P(\omega) = \sigma(\omega)\frac{\omega^3}{8 \pi^3}\frac{1}{\exp(\omega/k_B T)-1} \,,
\end{equation}
where $\sigma(\omega)$ is known as a \textit{greybody factor}. One finds that this greybody factor is precisely the absorption cross-section for the appropriate particle incident on the black hole. In particular, if the black hole is to be in thermal equilibrium with a bath of particles surrounding it, the rate at which it emits these particles must equal the rate at which particles in the bath fall into the black hole. The relation between the absorption cross-section $\sigma$ and the absorption probability $\mathcal{A}$ is given by the usual relation
\begin{equation}
\label{optical}
\sigma = \frac{\pi}{k^2}\sum_\ell(2 \ell + 1 ) \mathcal{A}(\ell)  \,,
\end{equation}
where $\ell$ is the partial wave number and $k$ is the particle's momentum, given by $k = \sqrt{\omega^2 - m^2}$. We will calculate the quantity $\mathcal{A}$ by solving the field equations for the relevant particle in the black hole background, imposing purely ingoing boundary conditions at the horizon, corresponding to total absorption.

We will be interested exclusively in the low-momentum regime,  $k M \ll 1$. This corresponds to the wavelength of the particle being much larger than the Schwarzschild radius of the black hole.  Analytic results for both massless and massive particles are well-known in this regime. Furthermore, the greybody factor in the low-momentum regime will be determined by the $\ell = 0$ partial wave, for which there is a great simplification of the tidal terms in the action, seen in Eq. \eqref{simple}.

An analytic expression for the greybody factor in the regime where the de Broglie and Compton wavelengths of the particle are much larger than the Schwarzschild radius of the black hole (that is, $\omega M \ll 1$) was found in \cite{Unruh}:
\begin{equation}
\label{unruh}
\sigma(k) = 32 \pi^2 M^3\frac{ (2k^2 + m^2)\sqrt{k^2 + m^2}}{ k^2}\left(1 - \exp\left(-2 \pi M \frac{2k^2 + m^2}{k}\right) \right)^{-1} \,.
\end{equation}
For low-momenta massive particles, this becomes
\begin{equation}
\sigma(k) \simeq \frac{32 \pi^2 M^3 m^3}{k^2} \,.
\end{equation}
In the low-energy massless case, the limiting behaviour is rather different. Taking the $m=0$, $k \to 0$ limit of \eqref{unruh}, we have
\begin{equation}
\label{surface}
\sigma(k) \simeq 16 \pi M^2 \,.
\end{equation}

\subsection{Effective Field Theory of Extended Objects}

Recall that the action for a radially infalling extended object can be written 
\begin{equation}
S = -\int \mathrm{d} \tau \left(m + c \frac{6M^2}{r^6}\right)\,.
\end{equation}
Though the action in principle contains four powers of the velocity, for radial motion these appear precisely in the  combination of the 4-velocity to the fourth power (for more general motion this is not the case). We note that this amounts to the action for a relativistic point-particle, but with a position-dependent \textit{effective mass}
\begin{equation}
\mu(r) = m + c\frac{6M^2}{r^6} \,.
\end{equation}
The corresponding classical field theory that describes spinless coherent collections of these particles is thus the Klein-Gordon theory with such an effective mass:
\begin{equation}
\mathcal{L} = -\frac{1}{2}(\partial_\mu \phi)(\partial^\mu \phi) - \frac{1}{2}\mu(r)^2\phi^2 \,. \label{eft}
\end{equation}
We can formalise this correspondence as follows. First note that we can introduce an einbein $e(x)$ into the action to modify it thus\footnote{Indeed, this is the more appropriate action one should use for massless particles.}:
\begin{equation}
S = \frac{1}{2}\int \mathrm{d} \sigma \left( e^{-1} g_{\mu \nu}\dot{x}^\mu \dot{x}^\nu- e\mu(r)^2 \right) \,.
\end{equation}
Upon quantising this theory of a point-particle we find the effective action \cite{worldline1,worldline2}
\begin{equation}
\Gamma[g_{\mu \nu}] = \frac{1}{2} \mathrm{Tr} \ln(\nabla^2 - \mu(r)^2) \,,
\end{equation}
which is precisely the effective action of the Klein-Gordon theory \eqref{eft}. The classical equation of motion derived from this Lagrangian is simply
\begin{equation}
g^{\mu \nu}\nabla_\mu \nabla_\nu \phi - \mu(r)^2 \phi = 0 \,. \label{eom}
\end{equation}
\subsection{Solving The Equations of Motion}
It remains to solve equation \eqref{eom} subject to appropriate boundary conditions. We will restrict our analysis to the $\ell = 0$ partial wave solution of this equation, since the equation is derived from the assumption of radial particle motion. In Schwarzschild coordinates (now using $t$ to mean Schwarzschild time), we decompose our field $\phi$ as 
\begin{equation}
\phi(x) = e^{- i \omega t} R(r) \,.
\end{equation}
Then equation \eqref{eom} reads
\begin{align}
  \frac{1}{r^2}\left(1-\frac{2M}{r}\right)\partial_r \left(r^2\left(1-\frac{2M}{r}\right)\partial_r \right)R(r) - \mu(r)^2\left(1-\frac{2M}{r}\right)R(r) + \omega^2 R(r) = 0 \,. 
\end{align}
We can simplify this equation using the tortoise radial coordinate $r_*$:
\begin{equation}
r_* = r + 2M \ln\left(\frac{r}{2M}-1\right) \qquad \implies \qquad \diff{r_*}{r} = \left(1-\frac{2M}{r}\right)^{-1} \,.
\end{equation}
Then the equation of motion becomes, with $S = rR$ \cite{MTW},
\begin{equation}
\label{tortoise}
-\ddiff{S}{r_*} + V(r)S = \omega^2 S \qquad\text{with}\qquad V(r) = \left(\mu(r)^2 + \frac{2M}{r^3}\right)\left(1 - \frac{2M}{r}\right) \,,
\end{equation}
where $r$ is implicitly a function of $r_*$. To calculate the absorption probability, we impose purely \textit{ingoing} boundary conditions at the horizon. When $r = 2M$, this equation becomes that of a simple harmonic oscillator, with solution
\begin{equation}
R(r_*) = A_+ \exp(i \omega r_*) + A_- \exp(-i  \omega r_*) \,.
\end{equation}
Ingoing boundary conditions correspond to $A_+ = 0$. At large distances, the effects of the tidal terms are negligible. In the massless case, the solutions are spherical Bessel functions, whilst the long-range $2M/r$ term in the massive case yields Coulomb wavefunction solutions, which in the far-field region have the asymptotic form \cite{handbook}:
\begin{equation}
\label{ff}
R(r) = \frac{B_+}{r}\exp(i k r - i \eta \ln (2 k r) + i \theta) + \frac{B_-}{r} \exp(-i k r + i \eta \ln (2 k r) - i \theta) \,,
\end{equation}
where $\eta=- M(2k^2+m^2)/2k$ and $\theta$ is an unimportant constant phase.

Starting with boundary conditions at the horizon, we numerically integrate this equation out to large $r$ to extract the reflection coefficient
\begin{equation}
\mathcal{R} = \left|\frac{B_+}{B_-}\right|^2 \,.
\end{equation}
The absorption probability we desire is then simply
\begin{equation}
\mathcal{A} = 1 - \mathcal{R} \,,
\end{equation}
and from here the greybody factor can be computed according to Eq. \eqref{optical}.

\subsection{Results}

We present in this section the results of the numerical simulations. We first consider the regime where the mass and energy of the particle are small, in the sense that $\omega M \ll 1$ and $m M \ll 1$. Thereafter we consider the large mass ($mM \gg 1$) case.

\subsubsection{Low Mass Emission}

\begin{figure}[H]
\centering
    \makebox[\textwidth]{\makebox[1.1\textwidth]{%
    \begin{minipage}{.55\textwidth}
        \centering
        \includegraphics[width=1\textwidth]{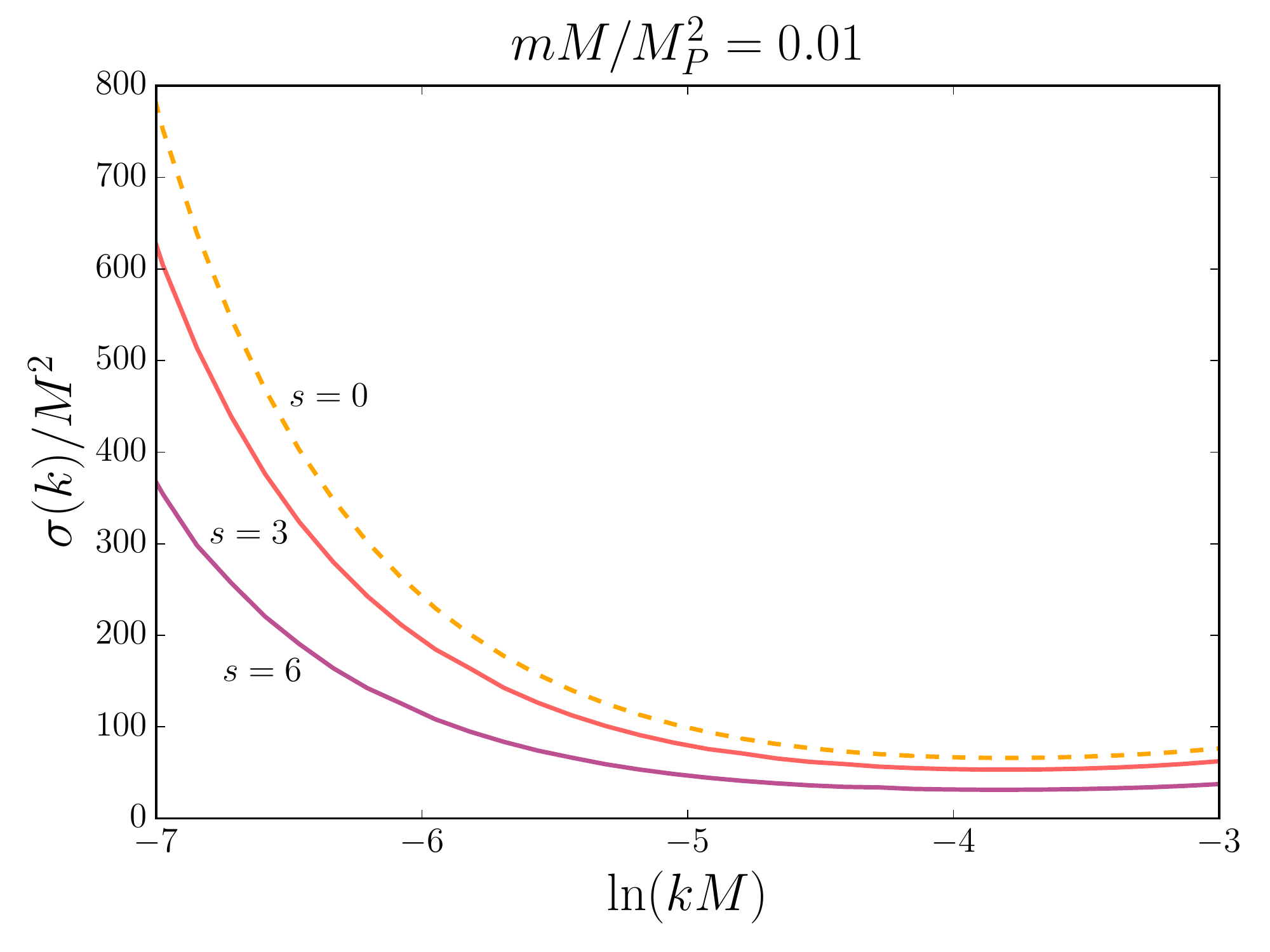}
    \end{minipage}\hfill
    \begin{minipage}{.55\textwidth}
        \centering
        \includegraphics[width=1\textwidth]{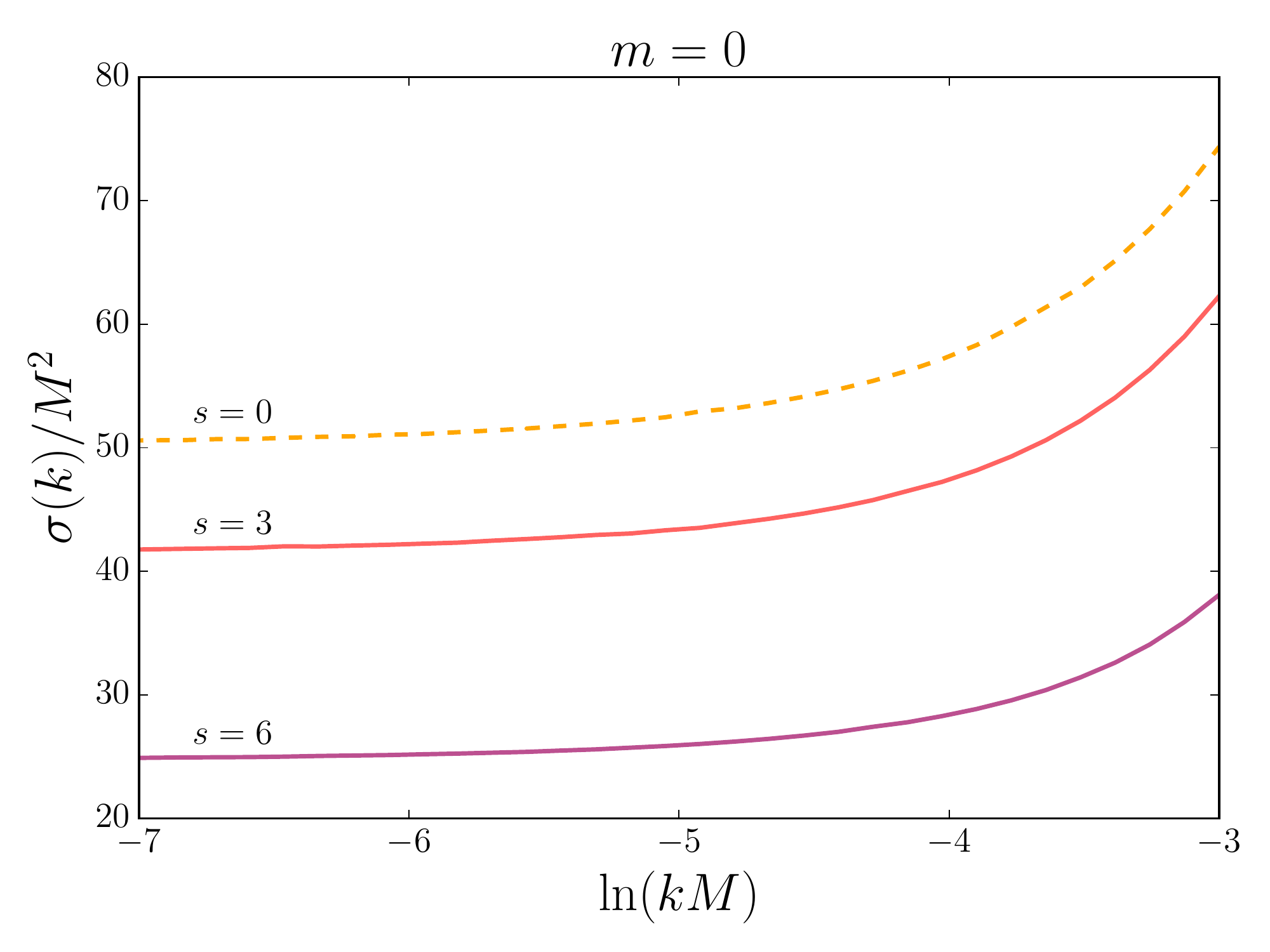}
    \end{minipage}}}
    \caption{Cross-section $\sigma$ against momentum $k$, for three choices of $s = c/M^3$. On the left is $\sigma$ in the low-mass regime, and on the right is the massless cross-section. One sees that the tidal terms reduce the value of the cross-section by a factor independent of momentum $k$. One also observes the $1/k^2$ asymptotic dependence of the massive cross-section, in contrast to the constant limiting value of the massless cross-section.}
    \label{crosssections} \end{figure}

The effect of the finite size of the object is always to reduce the greybody factor $\sigma$. We plot in Figure \ref{crosssections} the behaviour of the greybody factor as a function of momentum $k$, in both the massive and massless case. In particular, we find that, irrespective of the wavelength of the particle, the effect of the tidal terms is to reduce the absorption probability by a constant factor which we call $X(c)$:
\begin{equation}
X(c) = \frac{\sigma(c)}{\sigma(0)} \,.
\end{equation}

\begin{figure}[h]
\centering
\includegraphics[width=0.75\textwidth]{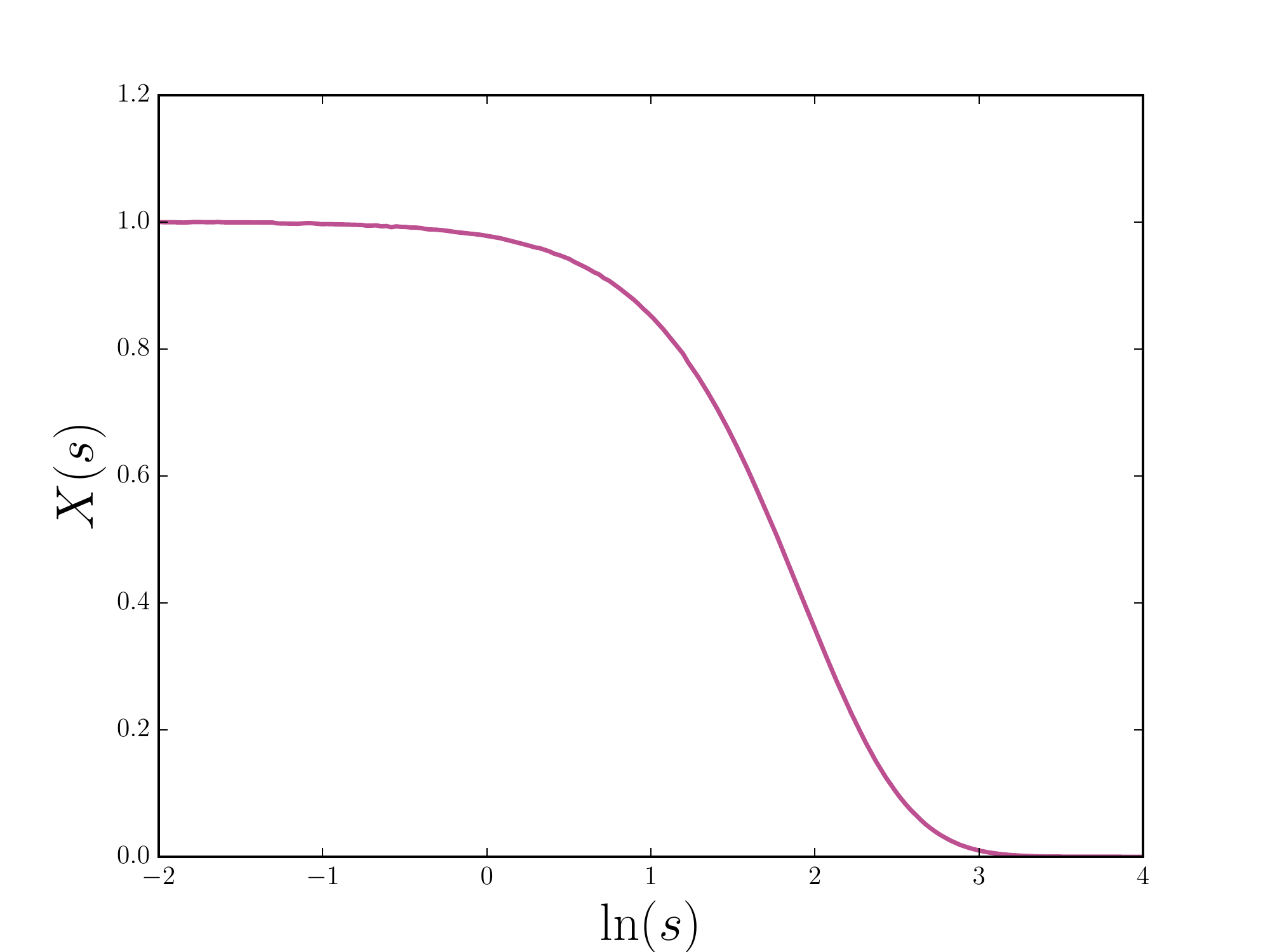}
\caption{The absorption cross-section (normalised to the point-particle cross-section) $X(s)$ against $s$, the strength of the tidal terms. Independent of the wavelength of the particle, the cross-section is reduced by a factor which becomes smaller as $s$ increases, reaches $1/2$ at around $s_{1/2}\simeq 5.9$, and eventually becomes exponentially small.}
\label{X}
\end{figure}

We find that for small $c$, $X(c)$ is equal to one, as expected, whilst for large $c$, $X(c)$ tends to zero, suppressing the greybody factor:
\begin{align}
X(0) = 1 \,,\\
\lim_{c\to \infty}X(c) = 0 \,.
\end{align}
We wish to know how this quantity behaves as a function of the parameters $m$ and $M$. We find that as a function of the dimensionless ratio
\begin{equation}
\label{s}
s = \frac{c}{M^3} \,,
\end{equation}
the quantity $X(s)$ is independent of the masses $m$ and $M$. The form of $X(s)$ as a function of $s$ is plotted in Figure \ref{X}. 

\begin{figure}[!]
\centering
    \makebox[\textwidth]{\makebox[1.1\textwidth]{%
    \begin{minipage}{.55\textwidth}
        \centering
        \includegraphics[width=1\textwidth]{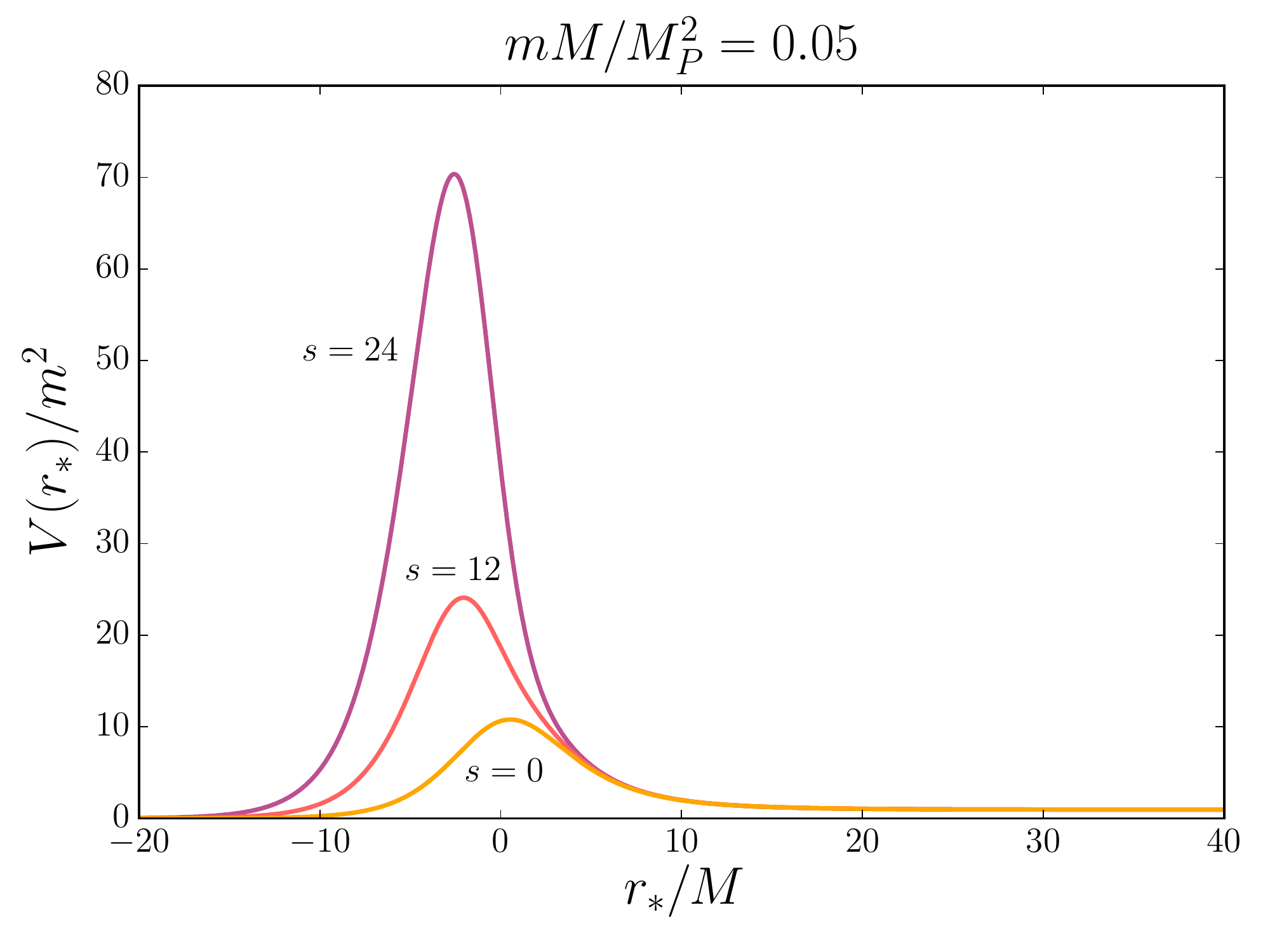}
    \end{minipage}\hfill
    \begin{minipage}{.55\textwidth}
        \centering
        \includegraphics[width=1\textwidth]{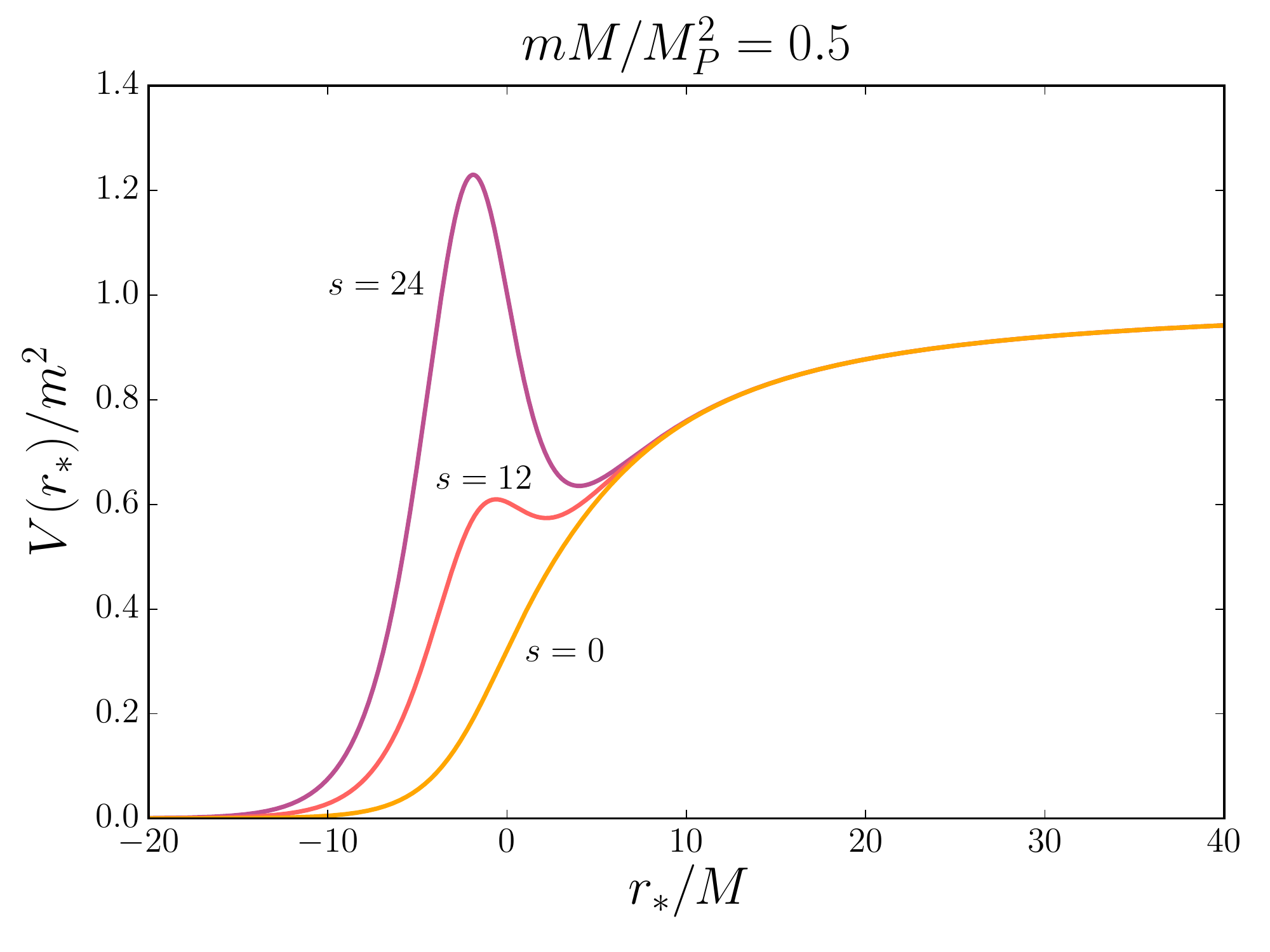}
    \end{minipage}}}
    \caption{The effective potential $V(r_*)$ from Eq. \eqref{tortoise} against $r_*$, for three choices of $s$, characterising the strength of the tidal term. On the left is this potential in the low-mass regime, and on the right with a mass close to the critical mass $m_\mathrm{crit} \simeq 0.385/M$. The effect of the tidal terms is to either produce or enhance a potential barrier through which the particle must tunnel, thereby reducing the probability of emission. }
    \label{potentials}
\end{figure}

The point at which the greybody factor is reduced to half, $s_{1/2}$, is given by 
\begin{equation}
 s_{1/2}\simeq 5.9 \,.
\end{equation}
We find that beyond this point, the tail of the function $X(s)$ is well-modelled by an exponential. Numerically we determine that
\begin{equation}
X(s) \sim \exp(-s/ s_0) \qquad \text{with} \qquad s_0 \simeq 3.3 \,.
\end{equation}

We can understand this scaling relation by rescaling Eq. \eqref{eom} in terms of a dimensionless radius $\rho = r/M$ (and its corresponding tortoise coordinate). We have
\begin{equation}
-\ddiff{S}{\rho_*} +  \left(\left(mM + \frac{6c}{M^3\rho^6}\right)^2 + \frac{2}{\rho^3}\right)\left(1 - \frac{2}{\rho}\right)S = \omega^2 M^2 S  \,.
\end{equation}
The dimensionless parameters in this equation are $mM$, $\omega M$ and $c/M^3$. The effective mass is a rapidly decaying function of $\rho$, and so we expect the short distance behaviour to depend only on $c/M^3$, and the large distance behaviour on $m$ and $\omega$. In principle, these parameters could be mixed together in a complicated way when we match in the intermediate regime. However, for small masses and energies, the absorption probability is expected to be very small (cf. Eqs. \eqref{optical}, \eqref{unruh} and \eqref{surface}). Equivalently, the coefficients $B_+$ and $B_-$ in \eqref{ff} are almost equal and opposite. The matching of the two solutions in the intermediate regime thus amounts to a single continuity condition, relating the size of the small $r$ solution (which depends only on $c/M^3$) and the size of the large $r$ solution (which depends only on $mM$ and $\omega M$). This explains the factorisation of the cross-section into the product of the point-particle cross-section and the tidal effects.

We can also understand this relation another way, examining the effective potential that the particle moves in. Eq. \eqref{tortoise} resembles the 1D Schr\"{o}dinger equation for a particle moving in a potential well. For the particle at infinity to reach the horizon, it must tunnel through a potential barrier, which exists for all values of $c$ for sufficiently light particles. We plot this effective potential for certain values of $s$ in Figure \ref{potentials}. We hence expect the absorption probability to have an exponential suppression of the form
\begin{equation}
\label{WKB}
\mathcal{A} \sim \exp\left(-2 \int \mathrm{d}r_* \sqrt{V(r;c)-\omega^2}\right) \equiv\exp\left(-2 I(c)\right) \,.
\end{equation}
For small $m$, $\omega$, we find
\begin{equation}
I(c) \simeq \int_2^\infty \mathrm{d}\rho \sqrt{\left(\left(\frac{6c}{M^3\rho^6}\right)^2 + \frac{2}{\rho^3}\right)\left(1 - \frac{2}{\rho}\right)^{-1}} \,.
\end{equation}
We see again that this integral depends on $c$ only  through the dimensionless parameter $c/M^3$. Furthermore, numerical evaluation of this integral as a function of $c$ shows that it remains roughly constant until around $c/M^3 \sim 5$, and then rises approximately linearly, with gradient $1/3.3=1/s_0$. This explains the exponential suppression of the absorption probability. We plot this integral against $s$ on the left of Figure \ref{int}.


\subsubsection{High Mass Emission}

In the high-mass regime, the restriction $kM \ll 1$ corresponds to consideration of only non-relativistic particles. We can understand the behaviour of the absorption probability in this regime by considering once again the behaviour of the effective potential. It was shown in \cite{highmass} that for $m$ greater than a critical value $m_\mathrm{crit} \simeq 0.385/M$, the effective potential becomes a monotonically increasing function of $r$. Above this mass, therefore, there should be a large enhancement of the absorption cross-section due to the disappearance of a barrier to tunnel through. However, for a sufficiently large tidal term, the potential once more develops a barrier, and we expect exponential suppression of the cross-section. 

\begin{figure}[h]
\centering
    \makebox[\textwidth]{\makebox[1.1\textwidth]{%
    \begin{minipage}{.55\textwidth}
        \centering
        \includegraphics[width=1\textwidth]{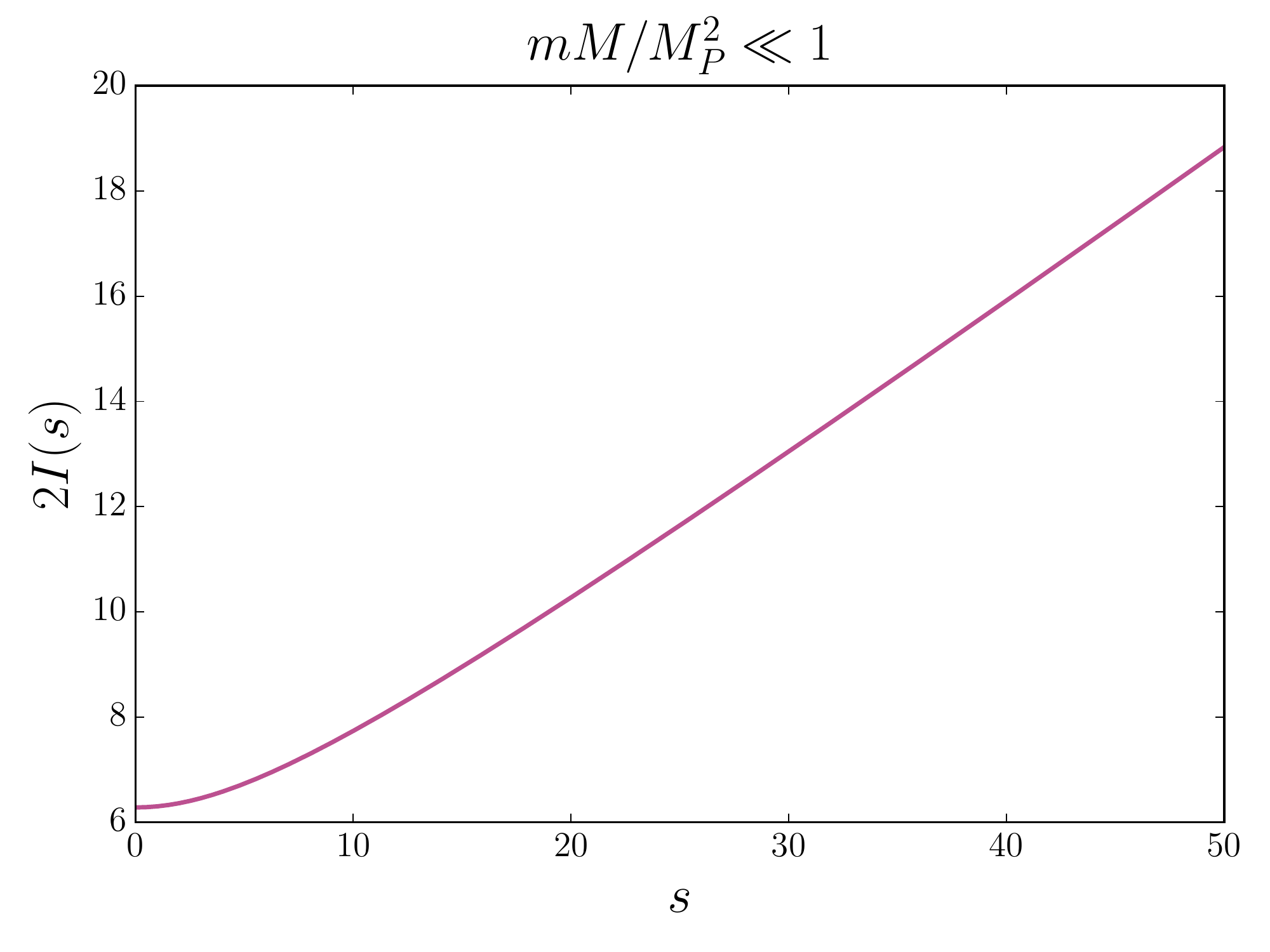}
    \end{minipage}\hfill
    \begin{minipage}{.55\textwidth}
        \centering
        \includegraphics[width=1\textwidth]{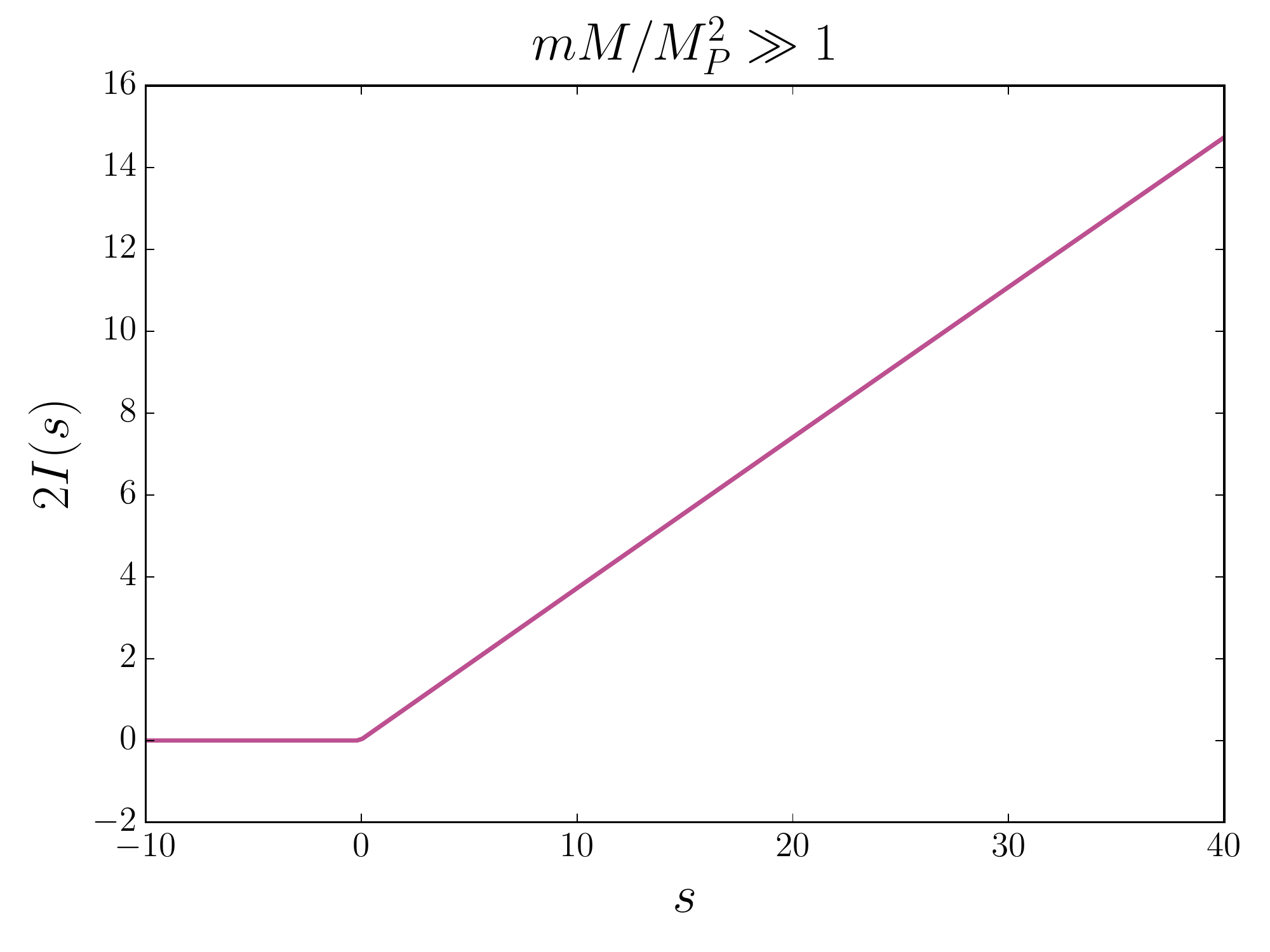}
    \end{minipage}}}
\caption{The barrier integral $I(s)$ defined by Eq. \eqref{WKB} against $s$. On the left is this function in the low-mass regime, and on the right the high-mass regime. For large $s$ the behaviour is linear, indicating an exponential dependence of the cross-section on $s$. In the high-mass regime, this linear behaviour only starts above a critical value $s_{1/2}$, itself linearly proportional to the particle mass $m$.}
\label{int}
\end{figure}

For large $m$ we find a simple analytic relation for the value of $s=c/M^3$ at which this barrier develops:
\begin{equation}
s_{1/2} \simeq \frac{1}{2}\left(\frac{3}{2}\right)^{11} m M \,.
\end{equation}
This relation is remarkably well borne out by numerical simulations for $mM \gtrsim 1$. We can again examine the integral $I(c)$ of our effective potential in the regime where $s$ is greater than $s_{1/2}$, and a barrier exists. As before, the integral depends linearly on the value of $s$, and so we expect exponential suppression of the cross-section for $s$ above $s_{1/2}$. The gradient is approximately $1/3.3$ as before, independent of the masses $m$ and $M$. Thus 
\begin{equation}
X(s) \sim \exp((s-s_{1/2})/s_0) \qquad \text{with} \qquad s_0 \simeq 3.3 \,.
\end{equation}
We plot the barrier integral against $s$ on the right of Figure \ref{int}.

\section{Discussion}

We have found that Hawking radiation of spatially extended objects always occurs at a lower rate to that of the corresponding pointlike objects, at least for scalar particles in the low-momentum regime. It is interesting to ask whether this effect is significant for extended objects in nature, such as protons, strings, GUT monopoles, or other objects. To estimate this, we need to find some relation between the physical size of these objects and the corresponding parameter $c$ in the action.

The parameter $c$ has dimensions of volume. For a given object, there are three characteristic length scales: its Compton radius $1/m$, its gravitational radius $m/M_P^2$, and its physical size $d$. For highly gravitationally-bound objects (we can consider the possibility of black holes radiating black holes themselves), the physical size is on the order of the gravitational radius, and a matching calculation indicates that $c \sim M_P^2 d^5$ \cite{houches}. For non-gravitational objects, we expect $d$ and $1/m$ to be the only relevant scales. However, we also expect the effects of the tidal forces to be well-behaved in the massless limit\footnote{For many extended objects, such as protons and GUT monopoles, the Compton and physical radii are comparable, so this argument is unnecessary.}, implying no dependence on the Compton wavelength: $c \sim d^3$.

We compute in the following table the value of the black hole mass $M$ (in Planck units) that results in an order two-fold suppression of the emission rate for various extended objects. The size and mass of the Q-ball are typical values taken from \cite{qball}.

\begin{center}
\begin{tabular}{lllr}
\hline
Object  & Mass $m$ & Size $d$ & $M/M_P$ \\
\hline
Fundamental String & 0 & $10^{-35}$ m & 1 \\
Proton & $1$ GeV & $10^{-15}$ m & $10^{20}$ \\
Q-ball & $10^{8}$ GeV &  $10^{-17}$ m& $10^{16}$ \\
GUT Monopole & $10^{16}$ GeV & $10^{-31}$ m & $10^4$ \\
Micro BH & $m$ & $m/M_P^2$  & $m/M_P$ \\
\hline
\end{tabular}
\end{center}

Finally, we must discuss the limitations of this analysis. Firstly, by reducing the theory of extended objects to an effective theory involving couplings to the curvature, we implicitly assume the object is rigid. For non-rigid objects, the point-particle approximation is inappropriate. However, this analysis should be valid whenever there is a sufficiently large mass gap in the spectrum of the extended object. For sufficiently low energies, we can expect the particle to remain in its ground state, and hence behave as a rigid object. We should also expect that for very large curvatures, it is not sufficient to take only the first term in the expansion of the action, and that higher-order terms become relevant.

Secondly, we have restricted attention to the long wavelength regime. In this regime the effects on the spectrum are dominated by $s$-wave radiation, for which there is a large simplification of the effective action. Furthermore, we assumed in our analysis that the emitted particles are spinless, which led us to an effective theory of a scalar field. We refer the analysis of higher spins and higher partial waves to future work.

Thirdly, we have only considered radiation from Schwarzschild black holes. It would be interesting to study how finite size effects modify the spectrum of rotating or charged black holes, or of black holes in different spacetime dimensions. For charged black holes radiating charged objects, for instance, there will be forces on the object as it escapes to infinity due to inhomogeneities in the electric field as well as those in the gravitational field, which would likely give rise to a rich range of phenomena.

\section*{Acknowledgements}
GJ is supported by the Science and Technology Facilities Council (STFC).

\bibliography{hawk}
\bibliographystyle{JHEP}
\end{document}